\documentclass[12pt]{article}

\usepackage{amsmath,amssymb,graphicx}

\makeatletter
\@addtoreset{equation}{section}
\makeatother

\newcommand{\beeq}{\begin{equation}}
\newcommand{\eneq}{\end{equation}}
\newcommand{\beqn}{\begin{eqnarray}}
\newcommand{\eeqn}{\end{eqnarray}}

\def\la{\raise.16ex\hbox{$\langle$}\lower.16ex\hbox{}  }
\def\ra{\raise.16ex\hbox{$\rangle$}\lower.16ex\hbox{} }
\def\go{\rightarrow}
\def\yield{\Longrightarrow}

\def\Tr{{\rm Tr \,}}

\def\eff{{\rm eff}}
\def\onehalf{ \hbox{$\frac{1}{2}$} }

\def\onethird{ \hbox{$\frac{1}{3}$} }
\def\twothird{ \hbox{$\frac{2}{3}$} }
\def\onequarter{ \hbox{$\frac{1}{4}$} }

\def\cL{{\cal L}}
\def\cG{{\cal G}}
\def\cH{{\cal H}}
\def\KK{{\rm KK}}
\def\psibar{ {\overline{\psi}} }

\begin{document}

\thispagestyle{empty}


\setcounter{section}{0}
\setcounter{equation}{0}
\setcounter{figure}{0}
\baselineskip 5mm
\rightline{\small OU-HET 925 (v2)}

\vspace{3.0cm}

\baselineskip=35pt plus 1pt minus 1pt
\begin{center}
{\Large \bf New dimensions from gauge-Higgs unification\footnote{To appear in the Proceedings of 
{\it Corfu Summer Institute 2016 ``School and Workshops on Elementary Particle Physics and Gravity"}, 
Corfu, Greece,  31 August - 23 September, 2016, and 
in the Proceedings of 
{\it ``6th CST-MISC Joint Symposium on Particle Physics -- from Spacetime Dynamics to Phenomenology --''}, Maskawa Institute for Science and Culture, Kyoto Sangyo
University, Japan, 15 - 16 October, 2016.}
}%
\end{center}

\vspace{1.5cm}
\baselineskip=22pt plus 1pt minus 1pt
\begin{center}
{\bf Yutaka Hosotani }
\vskip 5pt
{\small \it Department of Physics, Osaka University, Toyonaka, Osaka 560-0043, Japan}
\end{center}

\vskip 3.cm
\baselineskip=20pt plus 1pt minus 1pt
\begin{abstract}
The Higgs boson is unified with gauge fields in the gauge-Higgs unification.
The $SO(5) \times U(1)$ gauge-Higgs electroweak unification in the Randall-Sundrum 
warped space yields almost the same phenomenology at low energies as the standard model, 
and gives many predictions for the Higgs couplings and  new $W', Z'$ bosons around 
$6 \sim  8\,$TeV, which can be tested at 14$\,$TeV LHC.
The gauge-Higgs grand unification is achieved in $SO(11)$ gauge theory.
It  suggests the existence of the sixth dimension (GUT dimension) 
in addition to the fifth dimension (electroweak dimension).
The proton decay is naturally suppressed in the gauge-Higgs grand unification.
\end{abstract}

\newpage

\baselineskip=20pt plus 1pt minus 1pt
\parskip=0pt

\section{Standard model}

The Standard Model (SM) is very successful at low energies.
It is  gauge theory of $SU(3)_C \times SU(2)_L \times U(1)_Y$, 
whose Lagrangian density consists of four parts;
\beqn
\cL ~~~ &=& \cL_{\rm gauge} + \cL_{\rm Higgs} + \cL_{\rm fermion} 
+ \cL_{\rm Yukawa} ~, \cr
\noalign{\kern 10pt}
\cL_{\rm gauge} &=& - \onehalf \Tr G_{\mu\nu} G^{\mu\nu}
 - \onehalf \Tr F_{\mu\nu} F^{\mu\nu} - \onequarter B_{\mu\nu} B^{\mu\nu} ~, \cr
\noalign{\kern 10pt}
\cL_{\rm Higgs}  &=& |D_\mu \Phi |^2 - V[\Phi] ~, \cr
\noalign{\kern 10pt}
\cL_{\rm fermion} &=& \sum \psibar_j i \gamma^\mu D_\mu \psi_j ~, \cr
\noalign{\kern 10pt}
\cL_{\rm Yukawa} &=& \sum \big\{ y_{jk}^d \psibar_j \Phi \psi_k + 
y_{jk}^u \psibar_j \tilde \Phi \psi_k \big\}  + (h.c.) ~.
\label{SMaction1}
\eeqn
The  form of the parts $\cL_{\rm gauge}$ and $\cL_{\rm fermion}$ is 
determined by the gauge principle, and is beautiful.
The form of the Higgs potential $V[\Phi]$ in $\cL_{\rm Higgs}$, however, 
 is given in ad hoc manner.
The Yukawa couplings $y_{jk}^{u,d}$ in $\cL_{\rm Yukawa}$ are arbitrary as well.
The parts $\cL_{\rm Higgs}$ and $\cL_{\rm Yukawa}$ lack a principle.

The electroweak (EW) gauge symmetry breaking in the SM is brought about  
by an intentional  choice of $V[\Phi]$ which is assumed to have
a global minimum at $\la \Phi \ra \not= 0$.
In other words, the EW gauge symmetry breaking is enforced by hand.
The Higgs boson remains mysterious in the SM.

\section{Gauge-Higgs unification}

In the gauge-Higgs unification one starts with gauge theory in higher 
dimensions.\cite{YH1, Davies1, Hatanaka1998}
The Higgs field becomes a part of the extra-dimensional component of gauge 
fields.   Schematically
\beqn
\cL_{\rm gauge} + \cL_{\rm Higgs} &\yield& \cL_{\rm gauge}^{\rm 5d} ~, \cr
\noalign{\kern 10pt}
\cL_{\rm fermion} + \cL_{\rm Yukawa}  &\yield& \cL_{\rm fermion}^{\rm 5d} ~.
\label{GHaction1}
\eeqn
The effective Higgs potential is generated dynamically at the quantum level
from $\cL_{\rm gauge}^{\rm 5d} + \cL_{\rm fermion}^{\rm 5d}$.
In short, the theory  is governed by the gauge principle, and  
becomes concise and beautiful.\cite{HY2015, Hosotani2016b}

In the gauge-Higgs unification in five dimensions $(x^\mu, y)$
\beeq
A_M = 
\begin{cases} 
 A_\mu  \supset \gamma, W, Z \cr 
 \noalign{\kern 5pt}
 A_y \supset {\rm Higgs~ boson} \sim \hbox{Aharonov-Bohm (AB) phase}~ \theta_H
\end{cases}
\label{GHpattern1}
\eneq
When the fifth dimension is not simply connected, the Higgs field appears as
an Aharonov-Bohm phase $\theta_H$  in the fifth dimension.  The effective potential
$V_\eff (\theta_H)$ becomes nontrivial at the one-loop level.  
When $V_\eff (\theta_H)$ is minimized at $\theta_H \not= 0$, the 
EW symmetry is dynamically broken.  Finite Higgs boson mass is generated.  
The gauge-hierarchy problem is solved.

\section{$SO(5) \times U(1)$ gauge-Higgs EW unification}

The Randall-Sundrum (RS) warped space is specified with the metric
\beqn
&&\hskip -1.cm
ds^2 = e^{- 2\sigma (y)} dx^\mu dx_\mu + dy^2~, \cr
\noalign{\kern 10pt}
&&\hskip -1.cm
\sigma(y) = \sigma(-y) = \sigma(y+ 2L) ~, \cr
\noalign{\kern 10pt}
&&\hskip -1.cm
\sigma (y) = k |y| \quad {\rm for~} | y | \le L.
\label{RS1}
\eeqn
The RS space has  topology of $M^4 \times (S^1/Z_2)$, in which
$(x^\mu, y), (x^\mu, -y)$ and $(x^\mu, y+2L)$ are identified.
Its fundamental region  is 5d AdS space sandwiched by  
UV and IR branes, at $y_0 =0$ and $y_1=L$.
The 5d cosmological constant is given by $\Lambda = - 6 k^2$.
The $SO(5)$ and $U(1)_X$ gauge fields, $A_M$ and $B_M$, satisfy
\beqn
&&\hskip -1.cm
\begin{pmatrix} A_\mu \cr A_y \end{pmatrix} (x, y_j - y) =
P_j \begin{pmatrix} A_\mu \cr -A_y \end{pmatrix} (x, y_j + y) P_j^{-1} ~, ~~
P_j \in SO(5) ~, ~~ P_j^2 = 1 ~, \cr
\noalign{\kern 10pt}
&&\hskip -1.cm
\begin{pmatrix} B_\mu \cr B_y \end{pmatrix} (x, y_j - y) =
\begin{pmatrix} B_\mu \cr -B_y \end{pmatrix} (x, y_j + y) ~,  
\label{BC1}
\eeqn
Although gauge potentials themselves are not single-valued, 
physical gauge-invariant quantities are single-valued.\cite{ACP2005}-\cite{FHHOS2013} 

The set of the matrices $P_0, P_1$ is called the orbifold boundary condition.
We take
\beeq
P_0 = P_1 = \begin{pmatrix} -1 &&&&\cr & -1 &&& \cr && -1 &&\cr
&&&-1 & \cr &&&&+1 \end{pmatrix} ,
\label{BC2}
\eneq
by which gauge symmetry $\cG = SO(5) \times U(1)_X$ is reduced to
$\cH = SO(4) \times U(1)_X$.
Zero modes (parity even-even modes) appear in the $\cH$
part of $A_\mu, B_\mu$, and in the $\cG/ \cH$ part of $A_y$.
The latter is an $SO(4) \simeq SU(2)_L \times SU(2)_R$ vector,
or an $SU(2)_L$ doublet, corresponding to the 4d Higgs field in the SM.

Quark-lepton multiplets  are introduced in the vector representation of $SO(5)$
in the bulk.  In addition, one introduces dark fermions in the spinor representation
in the bulk.  On the UV brane at $y=0$ brane fermions in $SU(2)_L$ doublet and
a brane scalar $\Phi$ in $SU(2)_R$ doublet are introduced.
The brane scalar $\Phi$ spontaneously breaks $SU(2)_R \times U(1)_X$
to $U(1)_Y$, and at the same time gives rise  to additional mass terms for fermions.
The resultant spectrum at low energies ($< 1\,$TeV) is that of the SM.
The effective potential $V_\eff (\theta_H)$ is evaluated at the one loop.  
Contributions from the top quark multiplet and dark fermions triggers
the EW gauge symmetry breaking with a Higgs boson mass $m_H = 125\,$GeV.

\section{Success}

The $SO(5) \times U(1)_X$ gauge-Higgs unification  is successful.
The gauge principle governs the theory, including dynamics of the 4d 
Higgs boson.\cite{FHHOS2013}-\cite{FHHO2016}

\vskip 5pt
\noindent (1) 
The 4d Higgs boson, the four-dimensional fluctuation mode of
the AB phase $\theta_H$ in the fifth dimension, is massless at the tree level but
acquires a nonvanishing mass at the one loop level which is free from 
divergence and independent of regularization methods employed. 
The gauge hierarchy problem, a cumbersome problem in many theories, 
 is naturally solved.

\vskip 5pt
\noindent (2) 
The phenomenology at low energies ($\le 1\,$TeV) for $\theta_H < 0.1$ 
is almost the same as in the SM.   

\vskip 5pt
\noindent (3) 
There is no vacuum instability problem associated with the 4d Higgs scalar field.\cite{Degrassi2012}
The effective potential for the 4d Higgs field $H(x)$ is given by
$V_\eff (\theta_H + H/f_H)$.  The large gauge invariance guarantees the 
periodicity $V_\eff (\theta_H + 2 \pi)=V_\eff (\theta_H)$, which in turn implies 
that there never occurs the instability.  It has been explicitly shown that 
$V_\eff (\theta_H)$ is finite at the one loop level.

\vskip 5pt
\noindent (4) 
Dynamical EW symmetry breaking takes place in the RS space.   The existence of 
a heavy quark, the top quark $m_t > m_W$,  is crucial .  
$V_\eff (\theta_H)$ is controlled  by the $W$ and $Z$ bosons, the top quark
multiplet, and the dark fermions.  Light quarks and leptons multiplets are irrelevant
for the EW symmetry breaking in the RS space.

\section{Predictions}
The gauge-Higgs unification gives many predictions to be confirmed
by the forthcoming and future experiments.
Although the model contains several parameters, most of physical quantities
are determined by the AB phase $\theta_H$. 

\vskip 5pt
\noindent (a) 
The Yukawa couplings of quarks and leptons, $Y_\alpha$, the three-point
couplings of the Higgs boson to $W, Z$ bosons, $g_{HWW}, g_{HZZ}$, are
given, in good approximation, by\cite{HS2007, HK2008}
\beeq
Y_\alpha, g_{HWW}, g_{HZZ} 
\simeq (\hbox{SM values}) \times \cos \theta_H ~.
\label{Hcoupling1}
\eneq
The deviation from the SM is less than 1$\,$\% for $\theta_H < 0.1$.

\begin{figure}[bth]
\vskip -5.cm
\hskip 5.cm
\includegraphics[bb=171 346 862 518, width=10cm]{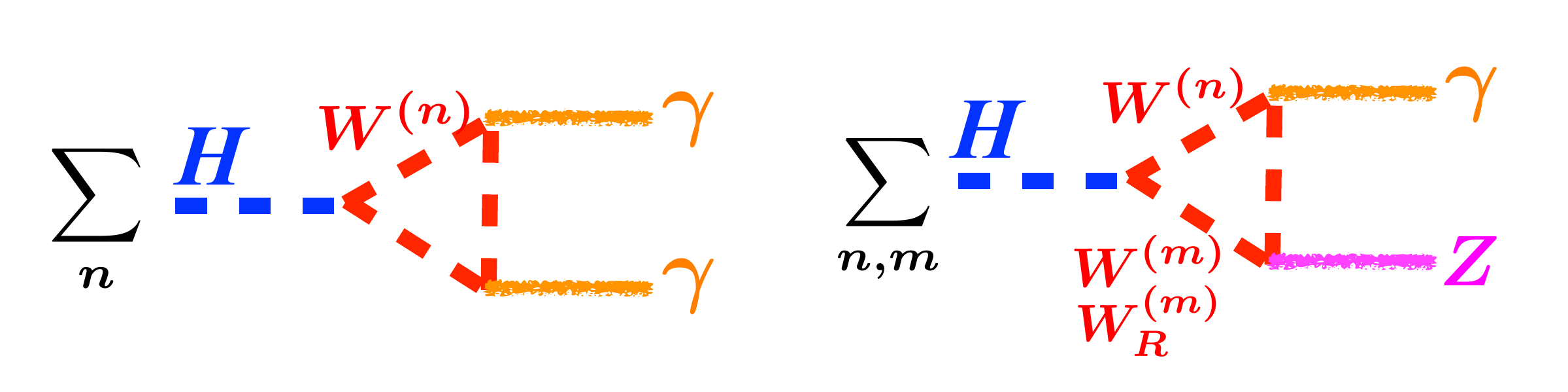}
\vskip 5.cm
\caption{Diagrams contributing to $H \go \gamma \gamma, Z \gamma$.  The infinite sums converge
and give small corrections to the SM.  There are diagrams in which the top quark and its KK tower
run inside the loops.}
\label{HiggsDecay1}
\end{figure}

\vskip 5pt
\noindent (b) 
Decay of the Higgs boson to  $\gamma \gamma$, $Z \gamma$, and 
two gluons take place through one-loop diagrams.  In the gauge-Higgs unification
an infinite number of various Kaluza-Klein (KK) modes run inside the loop.
(Fig.\ \ref{HiggsDecay1})
Each of their contributions gives $O(1/n)$ correction to the decay width
where $n$ is the KK number.  There appears miraculous cancellation among
them so that the sum of all contributions turns out finite and small.
It gives less than 1$\,$\% correction to those in which SM particles run inside 
the loop for $\theta_H < 0.1$.  The cancellation in the process $H \go Z \gamma$ is 
highly nontrivial, as the KK number can change inside the loop.\cite{FHHOS2013, FHH2015}

\vskip 5pt
\noindent (c) 
An immediate consequence of (a) and (b) is that both the production rate of the Higgs boson
at LHC and  decay widths of the Higgs boson to various modes are all suppressed 
in good approximation by a factor $\cos^2 \theta_H$ compared to those in the SM.  
Branching fractions of various decay modes are nearly the same as in the SM.
The signal strengths of the various decay modes are suppressed by a factor $\cos^2 \theta_H$.
For $\theta_H < 0.1$ the deviation from the SM is less than 1$\,$\%.

\begin{figure}
\vskip -3.5cm
\hskip 3.cm
\includegraphics[bb=78 223 909 508, width=13cm]{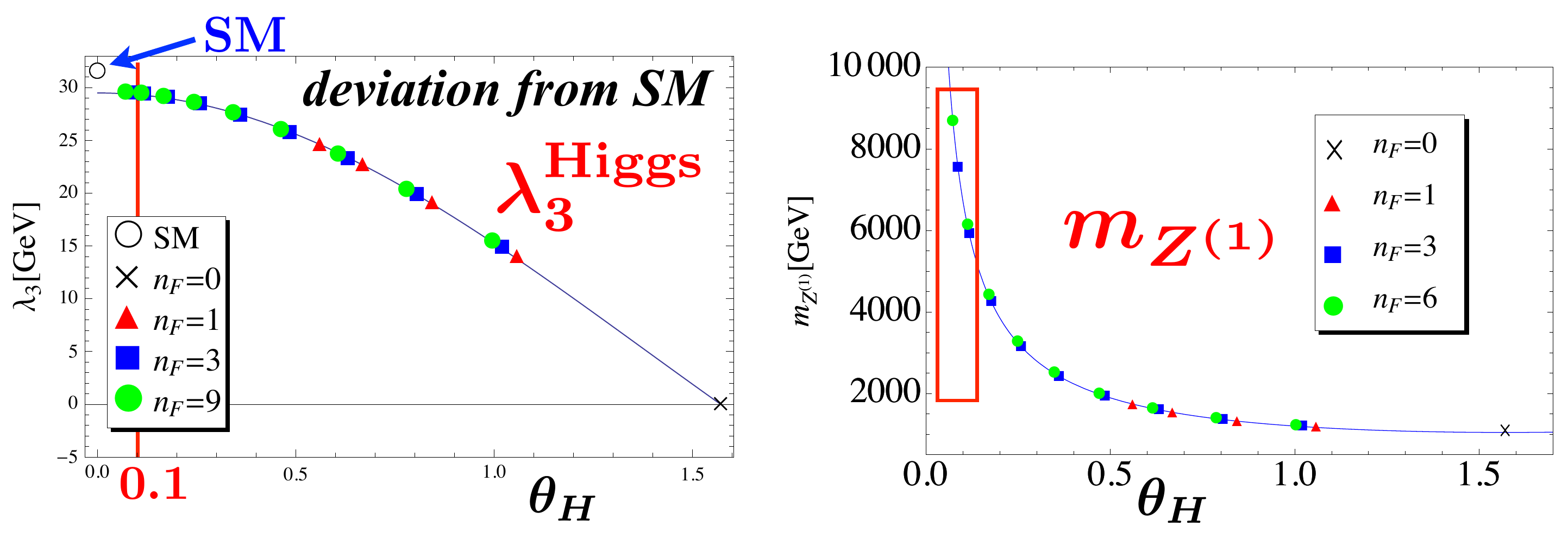}
\vskip 3.cm
\caption{Universality.  $\lambda_3^H$, $\lambda_4^H$, $m_{\rm KK}$, 
$m_{Z^{(1)}}$, $m_{\gamma^{(1)}}$, $m_{W^{(1)}}$ etc.\ are determined
by $\theta_H$, almost independent of the details of the model. 
In particular, they do not depend on the number $n_F$ of dark fermions.}
\label{universality1}
\end{figure}

\vskip 5pt
\noindent (d) 
The Higgs cubic and quartic self-couplings, $\lambda_3^H, \lambda_4^H$, deviate
from those in the SM, which can be tested in future.
Although the model has several parameters to be fixed, many of physical quantities
such as $\lambda_3^H$, $\lambda_4^H$, the KK mass scale $m_{\rm KK}$, and
the masses of the first KK modes $\gamma^{(1)}, Z^{(1)}, W^{(1)}$ depend 
only on $\theta_H$ in very good approximation.  This property is called the
universality.  (Fig.\ \ref{universality1})

The universality leads to strong prediction power in the gauge-Higgs unification.
Suppose that the first KK mode $Z^{(1)}$ is found at $m_0$.  
From the relation $ m_{Z^{(1)}} (\theta_H) = m_0$, the value $\theta_H$ is
determined.   Then other quantities $\lambda_3^H(\theta_H)$, 
$\lambda_4^H (\theta_H)$, $m_{W^{(1)}} (\theta_H)$ etc.\ are determined, 
and can be checked experimentally.

\begin{figure}
\vskip -5.cm
\hskip 2.7cm
\includegraphics[bb=112 324 955 592, width=14cm]{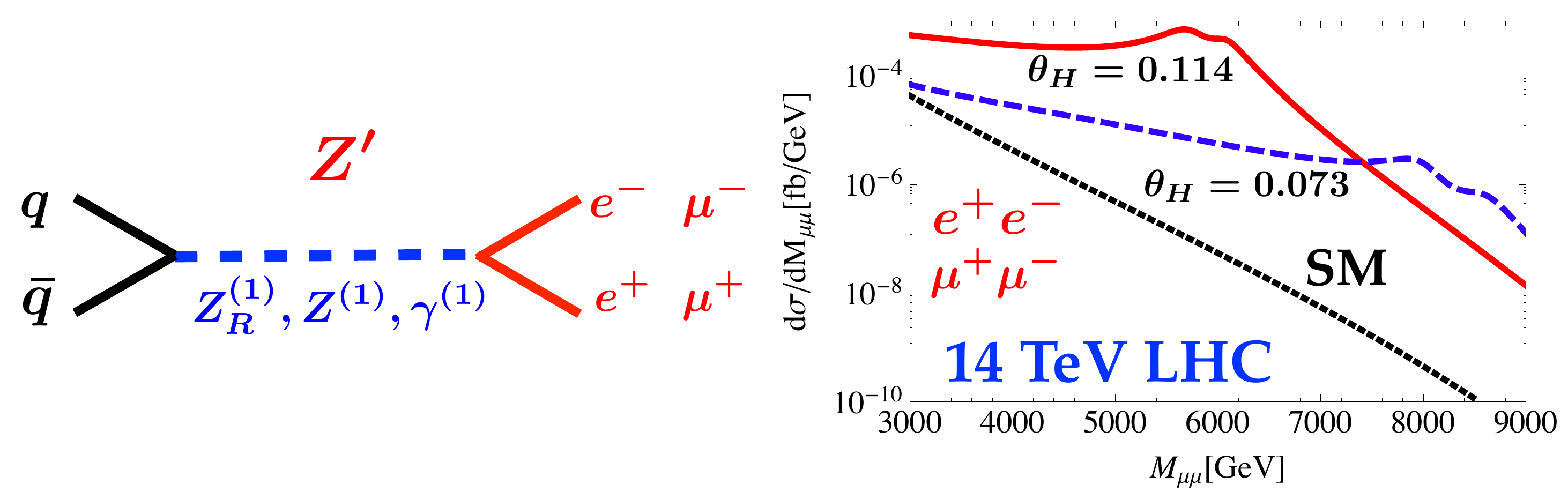}
\vskip 5.cm
\caption{$Z'$ production at LHC. }
\label{Zprime}
\end{figure}

\vskip 5pt
\noindent (e) 
The prediction of $Z'$ events gives the cleanest test of the model. (Fig.\ \ref{Zprime}) 
The first KK modes of the photon, $Z$ boson, and $Z_R$ boson appear as
$Z'$ events.  ($Z_R$ is associated with $SU(2)_R$, and has no zero mode.)
For $\theta_H = 0.114$, their masses are 
$(m_{Z_R^{(1)}}, m_{Z^{(1)}}, m_{\gamma^{(1)}}) = (5.73, 6.07, 6.08)\,$TeV 
and the widths are 
$(\Gamma_{Z_R^{(1)}}, \Gamma_{Z^{(1)}}, \Gamma_{\gamma^{(1)}})
= (482, 342, 886)\,$GeV.
For  $\theta_H = 0.073$, their masses are 
$(m_{Z_R^{(1)}}, m_{Z^{(1)}}, m_{\gamma^{(1)}}) = (8.00, 8.61, 8.61)\,$TeV 
and the widths are 
$(\Gamma_{Z_R^{(1)}}, \Gamma_{Z^{(1)}}, \Gamma_{\gamma^{(1)}})
= (553, 494, 1040)$ GeV.\cite{FHHO2016}

\section{$SO(11)$ gauge-Higgs grand unification}

It is necessary to incorporate strong interactions in the framework of gauge-Higgs
unification.  This leads to gauge-Higgs grand unification.\cite{Burdman2003}-\cite{Kojima2016}
We look for a scenario in which the EW Higgs boson appears as the extra-dimensional
component of gauge potentials, and electromagnetic, weak, and strong interactions
are unified in a single group, and no exotic particles appears at low energies.

One might think that the gauge group should contain $SO(5) \times U(1)_X \times SU(3)_C$
as a subgroup.  This turns out not to be  the case.  It is seen that $SO(11)$ gauge theory
does a job, keeping good features of the $SO(5) \times U(1)_X $ gauge-Higgs EW 
unification.\cite{HY2015a, Furui2016}

\begin{figure}[bth]
\vskip -3.5cm
\hskip 4.cm
\includegraphics[bb=124 262 905 518, width=11cm]{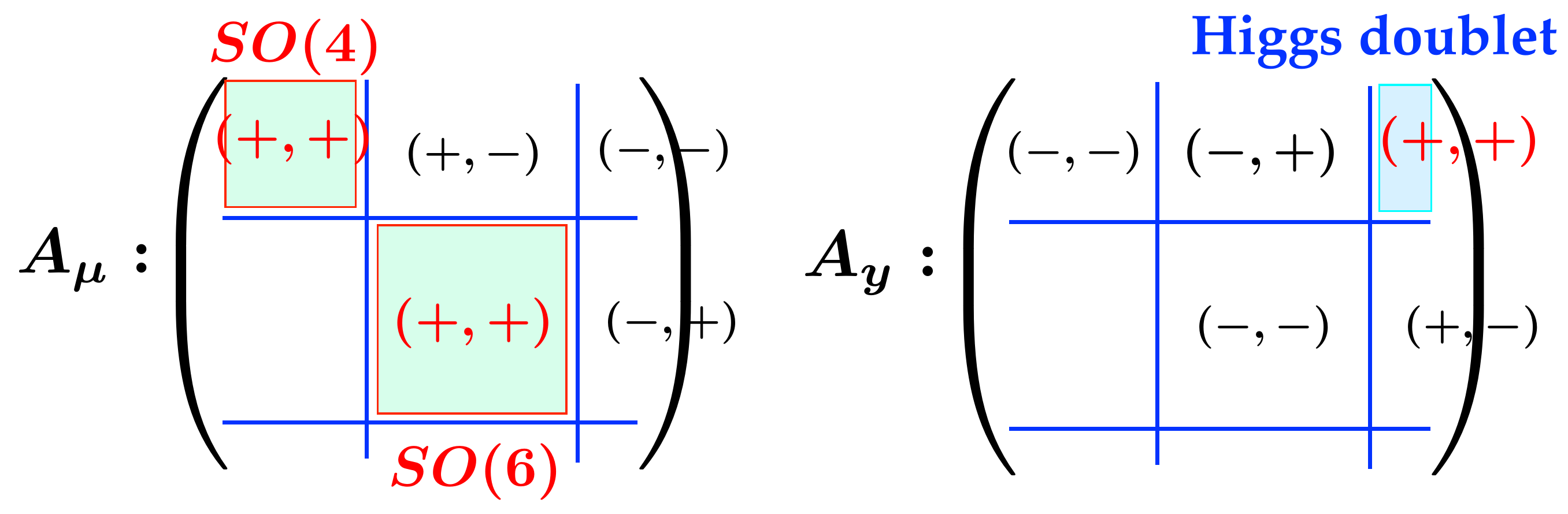}
\vskip 3.5cm
\caption{$SO(11)$ gauge-Higgs grand unification.  Parity $(P_0, P_1)=(+,+)$ modes
appear in the $SO(4) \times SO(6)$ block of $A_\mu$ and 
in the $[ SO(11)/SO(10) ] \cap  [ SO(11)/SO(4)\times SO(7)]$ block of $A_y$, 
$A_y^{a\, 11}$ ($a=1 \sim 4$).}
\label{GUT1}
\end{figure}

One starts with ${\cal G} = SO(11)$ gauge theory in the Randall-Sundrum warped space 
(\ref{RS1}).  The orbifold boundary condition is given by
\beqn
&&\hskip -1.cm
P_0^{\rm vec}=\mbox{diag}(I_{10},-I_1) ~,~~ P_1^{\rm vec} =\mbox{diag}(I_4,-I_7)~,  \cr
\noalign{\kern 5pt}
&&\hskip -1.cm
P_0^{\rm sp} = I_{16} \otimes \sigma^3 ~, \hskip 1.2cm
P_1^{\rm sp} = I_2 \otimes \sigma^3 \otimes I_8 
\label{grandBCP1}
\eeqn
in vectorial and spinorial representations.  
At the UV brane $SO(11)$ is broken to $SO(10)$ by $P_0$, whereas at the IR brane
it is broken to $SO(4) \times SO(7)$.
As a whole ${\cal G} = SO(11)$ is broken to ${\cal H} = SO(4) \times SO(6)$.  
Note that $SO(4) \simeq SU(2)_L \times SU(2)_R$, and $SO(6) \simeq SU(4)$.
At this stage $A_\mu$ has zero modes in the block ${\cal H}$.
On the other hand $A_y$ has zero modes in the block
$[ {\cal G}/SO(10) ] \cap  [ {\cal G}/SO(4)\times SO(7)]$.
In the vectorial representation $A_y$ has zero modes in the components
$A_y^{a \, 11}$ ($a= 1 \sim 4$), which correspond to the 4d Higgs field in the SM.
(Fig. \ref{GUT1})

\begin{figure}
\vskip -1.5cm
\hskip 3.cm
\includegraphics[bb=34 116 986 698, width=11cm]{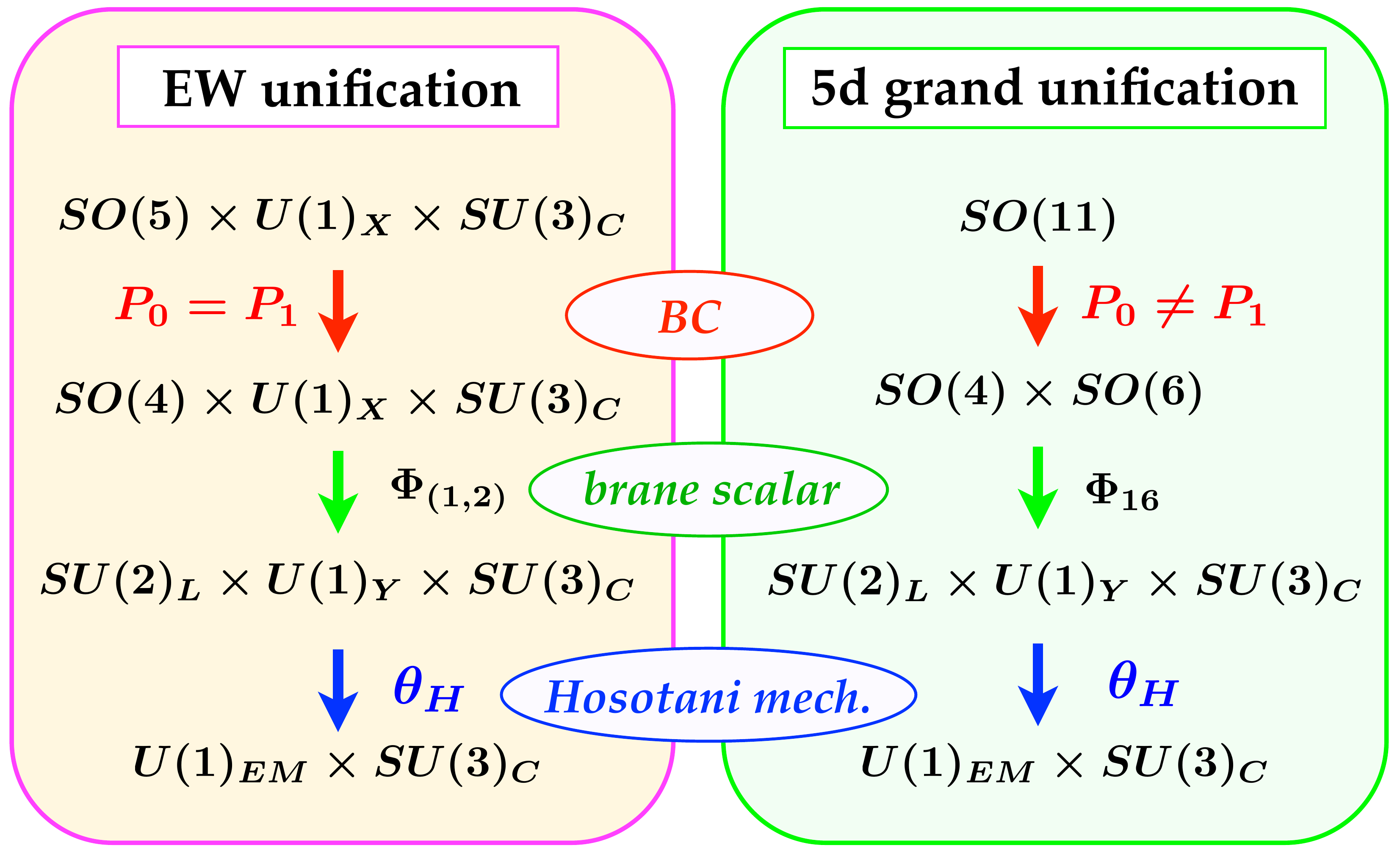}
\vskip 1.5cm
\caption{Comparison of gauge-Higgs EW and grand unification}
\label{GUT2}
\end{figure}

On the UV brane a brane scalar $\Phi_{\bf 16}$ is introduced.  $\Phi_{\bf 16}$
spontaneously breaks $SO(10)$ to $SU(5)$.  As a result ${\cal G} = SO(11)$ is
reduced to ${\cal G}_{\rm SM} = SU(2)_L \times U(1)_Y \times SU(3)_C$.
Note that $SU(3)_C \subset SO(6)$, and that  $U(1)_Y$ is a combination of
$SU(2)_R$ and $SO(6)$.  ${\cal G}_{\rm SM}$ is dynamically broken to
$U(1)_{\rm EM} \times SU(3)_C$ through the Hosotani mechanism.
The Weinberg angle at the GUT scale becomes $\sin^2 \theta_W = \frac{3}{8}$,
the same value as in the $SU(5)$ or $SO(10)$ GUT in four dimensions.
See the comparison of gauge-Higgs EW and grand unification in Fig.\ \ref{GUT2}.

Fermions are introduced  in the spinor ($\Psi_{\bf 32}$) and vector ($\Psi_{\bf 11}$) 
representations of $SO(11)$.  $\Psi_{\bf 32}$, for instance, satisfies
$\Psi_{\bf 32} (x,y_j - y)=  - \gamma^5 P_j^{\rm sp}  \Psi_{\bf 32} (x,y_j+y)$.
The content of $\Psi_{\bf 32}$ is given by
\beqn
&&\hskip -1.cm
\Psi_{\bf 32} = \begin{pmatrix} \Psi_{\bf 16} \cr \Psi_{\overline{\bf 16}} \end{pmatrix} ,~
\Psi_{{\bf 16}} =    \begin{pmatrix}
\nu \cr e \cr \noalign{\kern 1pt} 
\hat e \cr \hat \nu \cr  \noalign{\kern 1pt}
u_k \cr d_k \cr \noalign{\kern 3pt}
\hat d_k \cr \hat u_k 
\end{pmatrix} , ~
\Psi_{\overline{\bf 16}} =    \begin{pmatrix}
\nu' \cr e' \cr \noalign{\kern 1pt} 
\hat e' \cr \hat \nu' \cr  \noalign{\kern 1pt}
u_k' \cr d_k' \cr \noalign{\kern 3pt}
\hat d_k' \cr \hat u_k' 
\end{pmatrix} , 
~(k=1 \sim 3) ,\cr
\noalign{\kern 5pt}
&&\hskip -1.cm
\hbox{zero modes :} \quad 
\begin{pmatrix}
\nu_{L} \cr e_{L} 
\end{pmatrix}, ~
\begin{pmatrix}
u_{kL} \cr d_{kL} 
\end{pmatrix} ,~ 
\begin{pmatrix}
\nu_{R}' \cr e_{R}'
\end{pmatrix}, ~
\begin{pmatrix}
u_{kR}' \cr d_{kR}'
\end{pmatrix} . 
\label{F32content}
\eeqn
$\hat e$, $\hat u$, and $\hat d$ fields have
charges $+1$, $- \twothird$, and $+ \onethird$, respectively.
Zero modes appear only for the components of quarks and leptons.
Vector multiplets $\Psi_{\bf 11}$ are introduced to reproduce the mass spectrum
of down-type quarks and leptons.

One interesting feature is that all quarks and leptons appear in $\Psi_{\bf 32}$ as 
particles with the $\Psi$-fermion number $N_\Psi = +1$.
$N_\Psi$ is conserved even in the presence of $\Psi_{\bf 11}$.
A proton has $N_\Psi = 3$, whereas $\pi^0 e^+$ has $N_\Psi = -1$.
Thus the proton decay $ p \go \pi^0 e^+$ is forbidden.  
This should be contrasted to the situation in the 4d GUT.
In $SO(10)$ GUT in four dimensions a fermion multiplet is introduced  in the
spinor representation $\Psi_{\bf 16}$ for left-handed fields.
In the notation in (\ref{F32content}), $(u_k, d_k) \go (u_{kL}, d_{kL})$ and
$(\hat u_k, \hat d_k) \go ({u^c}_{kL}, {d^c}_{kL})$.  
As ${u^c}_{kL} \sim {u_{kR}}^\dagger$, gauge and Higgs interactions
convert a particle to an anti-particle, which induces proton decay.
In the gauge-Higgs grand unification such process is absent and 
the proton decay is naturally suppressed.

However, there is a problem.  Careful examination reveals that in the first and second
generations $\hat u, \hat d, \hat e$ have light masses, which contradicts the observation.
The source of this difficulty  lies in the fact that the parity at $y_0=0$ and $y_1 = L$ is
(even, odd) or (odd, even) for $\hat u, \hat d, \hat e$.  In the RS warped space
it leads to light masses.   In other words, $P_0 \not= P_1$ in the RS warped space
gives rise to a trouble.

\section{Gauge-Higgs grand unification in six dimensions}

\begin{figure}
\vskip -4.cm
\hskip 9.cm
\includegraphics[bb=296 295 696 517, width=6cm]{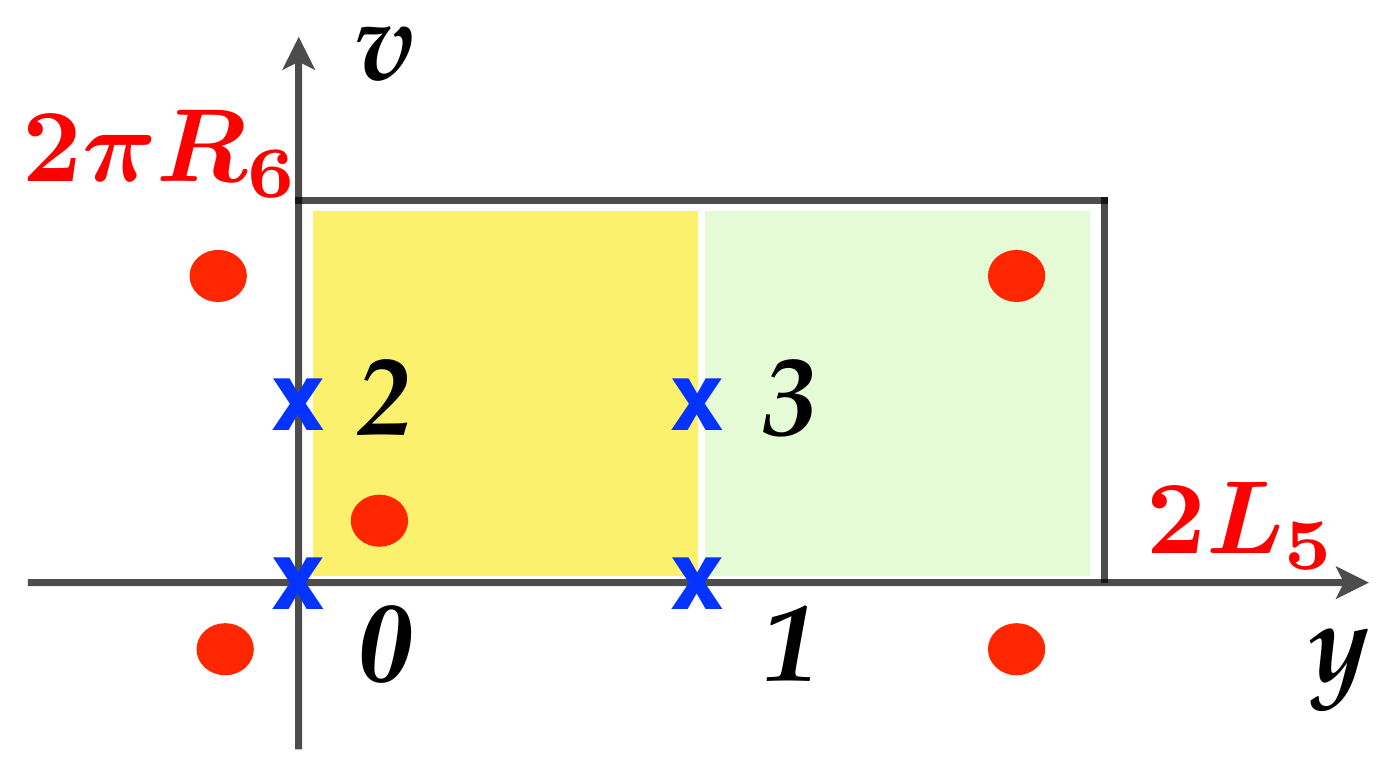}
\vskip 4.cm
\caption{Four fixed points (in blue) in the 6 dimensional gauge-Higgs grand unification.The fundamental region
is given by $0 \le y \le L_5, 0 \le v < 2 \pi R_6$.  
Red circle points represent a single spacetime point.
Around each fixed point, parity is defined.}
\label{GUT3}
\end{figure}

The difficulty is solved in gauge-Higgs unification in six-dimensional hybrid-warped
space.\cite{HY2017a}  Consider the six-dimensional space with a metric
\beqn
&&\hskip -1.cm
ds^2 = e^{- 2 \sigma (y)} ( dx^\mu dx_\mu + dv^2) + dy^2 ~, \cr
\noalign{\kern 10pt}
&&\hskip -1.cm
\sigma(y) = \sigma(-y) = \sigma(y+ 2L_5) ~,\cr
\noalign{\kern 10pt}
&&\hskip -1.cm
\sigma (y) = k |y| ~{\rm for~} | y | \le L_5.
\label{6dRS1}
\eeqn
We identify points
\begin{align}
(x^\mu, y, v) &\sim  (x^\mu, y + 2L_5, v) \sim (x^\mu, y, v+ 2\pi R_6) \cr
&\sim  (x^\mu, -y , -v)~.
\label{6dRS2}
\end{align}
The spacetime has topology of $M^4 \times (T^2/Z_2)$.
The fundamental region can be taken as $\{ 0 \le y \le L_5, 0 \le v < 2\pi R_6 \}$
The metric (\ref{6dRS1}) solves the Einstein equation with five-dimensional 
branes at $y=0$ and $y=L_5$. Six-dimensional spacetime is an AdS space with
$\Lambda = - 10 k^2$.  The sixth dimension is curled up in a circle with a very small
radius $R_6$.  We suppose that $z_L = e^{kL_5} \gg 1$ and 
\beeq
m_{\KK_5} = \frac{\pi k}{e^{k L_5} -1} \sim \pi k e^{- kL_5} 
\ll  m_{\KK_6} = \frac{1}{R_6} ~.
\label{6dRS3}
\eneq
Under $Z_2$ parity $(y,v) \go (-y, -v)$,  there appear four fixed points.  (See Fig. \ref{GUT3}.)
\beeq
(y_0, v_0) = (0,0) ,~ (y_1, v_1) = (L_5 , 0) ,~
(y_2, v_2) = (0, \pi R_6) ,~(y_3, v_3) = (L_5, \pi R_6) .
\label{6dRS4}
\eneq

We consider $SO(11)$ gauge theory in the 6-dimensional hybrid-warped space
(\ref{6dRS1}).  Gauge potentials $A_M$ satisfy
\beqn
&&\hskip -1.cm
\begin{pmatrix} A_\mu \cr A_y \cr A_v \end{pmatrix} (x, y_j - y, v_j - v) =
P_j \begin{pmatrix} A_\mu \cr -A_y  \cr - A_v  \end{pmatrix} (x, y_j + y, v_j + v ) P_j^{-1} ~, \cr
\noalign{\kern 10pt}
&&\hskip -1.cm
P_j ~  {\rm ~ or ~} - P_j  \in SO(11) ~, ~~ P_j^2 = 1 ~, ~~ P_3 = P_1 P_0 P_2 = P_2 P_0 P_1 ~.
\label{6dBC1}
\eeqn
Note that  only three of the four $P_j$'s  are independent, and
the condition $P_1 P_0 P_2 = P_2 P_0 P_1$ must be satisfied for the consistency.
We take, in place of (\ref{grandBCP1}), 
\beqn
&&\hskip -1.cm
P_0^{\rm vec}=P_1^{\rm vec} = \mbox{diag}(I_4,-I_7) ~, \cr
\noalign{\kern 10pt}
&&\hskip -1.cm
P_2^{\rm vec} = P_3^{\rm vec} = \mbox{diag}(I_{10},-I_1)~.
\label{grandBC2}
\eeqn

Fermion multiplets $\Psi_{\bf 32}$ and $\Psi_{\bf 11}$ are introduced in the bulk.
$\Psi_{\bf 32}$ is a 6d Weyl fermion, and satisfies
$\Psi_{\bf 32} (x, y_j - y, v_j -v) = P_j^{\rm sp} \bar \gamma 
\Psi_{\bf 32} (x, y_j + y, v_j +v)$ where $\bar \gamma = -i \Gamma^5 \Gamma^6$.
With this boundary condition zero modes appear chiral, with the quark-lepton content
given in (\ref{F32content}).  Furthermore,  the lightest modes of  hat  fields
$\hat e, \hat d, \hat u$ etc.\ have large masses of $O(R_6^{-1})$.

The symmetry breaking pattern is similar to the five-dimensional case.
The orbifold boundary condition in the sixth dimension reduces $SO(11)$ to $SO(10)$,
and the condition in the fifth dimension reduces $SO(11)$ to $SO(4) \times SO(7)$.
A brane scalar $\Phi_{\bf 32} (x,v)$ is introduced on the five-dimensional UV
brane at $y=0$.  It spontaneously breaks $SO(11)$ to $SU(5)$.
As a result the SM symmetry ${\cal G}_{\rm SM} =  SU(2)_L \times U(1)_Y \times SU(3)_C$ survives. 
By the Hosotani mechanism the symmetry is further broken to 
$U(1)_{\rm EM} \times SU(3)_C$.
Zero modes of $A_y$ correspond to the 4d Higgs doublet.  There appear
zero modes of $A_v$ in the same $SO(11)$ components as $A_y$,
which acquire masses of order $g   R_6^{-1}$ by the Hosotani mechanism.

\section{Summary}

The gauge-Higgs unification is promising. 
The $SO(5) \times U(1)$ gauge-Higgs EW unification gives definitive
predictions to be tested in the forthcoming LHC  experiments.
The incorporation of strong interactions leads to the $SO(11)$ gauge-Higgs grand
unification.  It seems necessary to introduce the sixth dimension to have
a spectrum consistent at low energies.  The fifth dimension serves as an EW 
dimension, whereas the sixth dimension as a GUT dimension.  
We are entering into an era of ``New Dimensions''.

\vskip .5cm

\noindent
{\bf Acknowledgement}

This work was supported in part by the Japan Society for the Promotion of Science, 
Grants-in-Aid for Scientific Research No 15K05052 .

\def\jnl#1#2#3#4{{#1}{\bf #2},  #3 (#4)}

\def\Zphys{{\em Z.\ Phys.} }
\def\jssc{{\em J.\ Solid State Chem.\ }}
\def\jpsJ{{\em J.\ Phys.\ Soc.\ Japan }}
\def\ptps{{\em Prog.\ Theoret.\ Phys.\ Suppl.\ }}
\def\PTP{{\em Prog.\ Theoret.\ Phys.\  }}
\def\PTEP{{\em Prog.\ Theoret.\ Exp.\  Phys.\  }}
\def\JMP{{\em J. Math.\ Phys.} }
\def\NPB{{\em Nucl.\ Phys.} B}
\def\NP{{\em Nucl.\ Phys.} }
\def\PLB{{\it Phys.\ Lett.} B}
\def\PL{{\em Phys.\ Lett.} }
\def\PRL{\em Phys.\ Rev.\ Lett. }
\def\PRB{{\em Phys.\ Rev.} B}
\def\PRD{{\em Phys.\ Rev.} D}
\def\PRe{{\em Phys.\ Rep.} }
\def\AP{{\em Ann.\ Phys.\ (N.Y.)} }
\def\RMP{{\em Rev.\ Mod.\ Phys.} }
\def\ZPC{{\em Z.\ Phys.} C}
\def\SCI{\em Science}
\def\CMP{\em Comm.\ Math.\ Phys. }
\def\MPLA{{\em Mod.\ Phys.\ Lett.} A}
\def\IJMPA{{\em Int.\ J.\ Mod.\ Phys.} A}
\def\IJMPB{{\em Int.\ J.\ Mod.\ Phys.} B}
\def\EPJC{{\em Eur.\ Phys.\ J.} C}
\def\PR{{\em Phys.\ Rev.} }
\def\JHEP{{\em JHEP} }
\def\JCAP{{\em JCAP} }
\def\cmp{{\em Com.\ Math.\ Phys.}}
\def\JPA{{\em J.\  Phys.} A}
\def\JPG{{\em J.\  Phys.} G}
\def\NJP{{\em New.\ J.\  Phys.} }
\def\PoS{{\em PoS} }
\def\CQG{\em Class.\ Quant.\ Grav. }
\def\ATMP{{\em Adv.\ Theoret.\ Math.\ Phys.} }
\def\ibid{{\em ibid.} }

\def\reftitle#1{{\it #1, }}    


\end{document}